\documentclass[11pt]{article}
\usepackage{arxiv} 
\usepackage{amsmath,amssymb}
\numberwithin{equation}{section}
\usepackage{subcaption}
\usepackage{graphicx}
\usepackage{booktabs}
\usepackage{siunitx}
\usepackage{hyperref}
\usepackage{algorithm}
\usepackage{algpseudocode}
\usepackage{listings}
\usepackage{xcolor}
\usepackage{tikz}
\usetikzlibrary{arrows.meta,positioning,calc}

\hypersetup{
  colorlinks=true,
  linkcolor=blue,
  citecolor=blue,
  urlcolor=blue
}
\lstdefinestyle{py}{
  language=Python,
  basicstyle=\ttfamily\small,
  keywordstyle=\color{blue},
  commentstyle=\color{gray},
  stringstyle=\color{teal},
  showstringspaces=false,
  breaklines=true,
  frame=single
}

\title{Real-time control of multiphase processes with learned operators}

\author{
 Paolo Guida \\
  Clean Energy Research Center (CERP)\\
  Physical Science and Engineering (PSE) Division \\
  King Abdullah University of Science and Technology\\
  Thuwal, Saudi Arabia 23955 \\
  \texttt{paolo.guida@kaust.edu.sa} \\
  \And
 Didier Barradas-Bautista \\
  Kaust Visualization Lab\\ 
  Core Lab Division\\
  King Abdullah University of Science and Technology\\
  Thuwal, Saudi Arabia 23955 \\
  \texttt{didier.bautista@kaust.edu.sa} \\
}

\begin{document}
\usetikzlibrary{arrows.meta,positioning,fit,calc,shapes.multipart}

\maketitle
\begin{abstract}
Multiphase flows frequently occur naturally and in manufactured devices. Controlling such phenomena is extremely challenging due to the strongly non-linear dynamics, rapid phase transitions, and the limited spatial and temporal resolution of available sensors, which can lead to significant inaccuracies in predicting and managing these flows. In most cases, numerical models are the only way to access high spatial and temporal resolution data to an extent that allows for fine control. While embedding numerical models in control algorithms could enable fine control of multiphase processes, the significant computational burden currently limits their practical application. This work proposes a surrogate-assisted model predictive control (MPC) framework for regulating multiphase processes using learned operators. A Fourier Neural Operator (FNO) is trained to forecast the spatiotemporal evolution of a phase-indicator field (the volume fraction) over a finite horizon from a short history of recent states and a candidate actuation signal. The neural operator surrogate is then iteratively called during the optimisation process to identify the optimal control variable. To illustrate the approach, we solve an optimal control problem (OCP) on a two-phase Eulerian bubble column. 
Here, the controller tracks piecewise-constant liquid level setpoints by adjusting the gas flow rate introduced into the system. The algorithm maps the predicted field to a relevant observable, the height of the liquid column in this case, and minimizes a receding-horizon objective. Because the resulting cost can be nonconvex and nonsmooth due to the level-extraction operation, which involves evaluating the volume fraction against a pre-established threshold, we employ Bayesian optimisation to select the inlet velocity that minimises the finite-horizon tracking error with few surrogate rollouts. The results we obtained indicate that field-level forecasting with FNOs are well suited for closed-loop optimization since they have relatively low evaluation cost. The latter provide a practical route toward MPC for fast multiphase unit operations and a foundation for future extensions to partial observability and physics-informed operator learning. \footnote{The code is available at \url{https://gitlab.com/paolo.guida/neuralmodelpredictivecontrol.git }}
\end{abstract}

\section{Introduction}\label{sec:introduction}
Multiphase flows are ubiquitous in engineering and characterise a large fraction of unit operations, biological systems, and emerging manufacturing technologies \cite{balachandar2010turbulent,worner2012numerical,yao2017application,lohse2022fundamental,manoharan2020manufacturing}. It is common to for processes to involve immiscible liquid-liquid dispersions in extraction equipments \cite{goodarzi2019comprehensive} where they govern droplet breakup/coalescence and interfacial mass transfer. In bio-processing and environmental engineering, multiphase reactors, such as bubble columns, are widely used due to their favourable hydrodynamics and enhanced mass transfer \cite{mcclure2015mixing}. Other gas-liquid systems include, for example, syngas production reactors via fermentation \cite{zhao2019gas} and spray cooling \cite{hu2017spray}. Beyond “canonical” process equipment, multiphase microfluidics is at play in inkjet-based additive manufacturing via jetting, breakup, wetting, and evaporation physics \cite{lohse2022fundamental}, while metal additive manufacturing involves highly transient vapour-liquid-solid interactions, such as keyhole instability and pore formation \cite{oh2011role}. Finally, multiphase phenomena also arise in electrochemical energy systems, where gas generation in lithium-ion batteries has significant safety implications \cite{liu2002treatment}, and even in the human body, where liquid-liquid separation occurs in cells, effectively creating a multiphase system \cite{alberti2019liquid}.
Many of the multiphase processes discussed above evolve over very short characteristic time scales, leaving little room for direct measurements, calculations and actuation, therefore limiting fine control in practice. 
Most control strategies for multiphase processes, therefore, do not involve feedback and are generally passive approaches. The latter, although often effective, may struggle when, for example, feed variability is significant.
It is also important to note that measurements are often not representative of the entire time evolution or the whole domain, as sensors are not always fast enough and cannot be placed arbitrarily in systems that experience high temperatures and pressures, for instance \cite{chaouki1997non,yang2006sensors,do2022capacitive}. Another challenge with multiphase flows is that some adopted sensors cannot be calibrated for both vapour and liquid, requiring \textit{ad hoc} calibration and the ability to distinguish between the two \cite{chaouki1997non, shi2020conductance}.
Model predictive control (MPC) is an attractive approach for handling fast transients and managing constraints \cite{aggelogiannaki2008nonlinear,grune2009analysis,arbabi2018data}. It is also worth mentioning the ability of such models to couple planning with constraints, as actuator limits, safety boundaries, and operational requirements can be effectively enforced while optimising performance over a finite horizon \cite{camacho2007constrained}.  However, multiphase systems are strongly nonlinear, often exhibiting regime transitions and hysteresis, and generally involve tightly coupled interfacial physics across multiple scales. As a result, constructing reduced-order models or accurate surrogates that capture all relevant underlying features of the problem is an exceptionally difficult task \cite{ganti2020data}. At the same time, high-fidelity CFD is typically far too computationally expensive for real-time prediction and optimization \cite{wu2017model}. Modelling such systems, in fact, often requires small internal time steps, repeated nonlinear/pressure-correction iterations, and interface reconstruction methods \cite{roenby2017new}. Embedding such solvers within the repeated rollouts required by MPC can therefore be challenging, particularly because updates must occur at the same time scale as sensing and actuation. While the observable of interest is often a low-dimensional variable, such as the level of liquid and/or the size distribution of an emulsion, that observable depends on the global evolution of coupled PDEs and boundary forcing. As a result, computational efficiency is of paramount importance to make MPC a practically implementable approach. 
A recent work by \cite{li2020neural} has introduced the concept of neural operators. Neural operators are a class of machine learning models that learn mappings between function spaces instead of finite-dimensional entities \cite{kovachki2021neural}. More in detail, while a general surrogate model like CNN or RNN learns a mapping $f:\mathbb{R}^n \rightarrow \mathbb{R}^m$, a neural operator learns a mapping between infinite spaces $\mathcal{G}:\mathcal{A} \rightarrow \mathcal{U}$ where $\mathcal{A}$ and $\mathcal{U}$ are spaces of functions. The latter have several advantages, most notably that they are theoretically discretisation-independent because input and output are functions \cite{kovachki2023neuraloperator, li2023geometry}, a feat that, while often mentioned and particularly appealing, presents some limitations in practice and is not what we exploit in this work. The characteristic we are interested in is, in fact, the ability to infer the evolution of complex systems extremely quickly, making neural operators a valid alternative to reduced-order models for real-time optimisation and control. 

Among NOs, Fourier Neural Operators (FNOs) are a class of neural operators that are particularly attractive for PDE systems, including the Navier-Stokes equations, since they represent global interaction well in the spectral domain (spectral methods are commonly used to solve the NS equations anyway) \cite{li2020fourier,kovachki2023neuraloperator}. We will therefore adopt FNOs despite the existence of other architectures that might be similarly effective, such as DeepONet or U-Net \cite{lu2021learning, ronneberger2015unet}. As mentioned earlier, for control, the key advantage is that FNO surrogates provide fast, differentiable multi-step forecasts, enabling repeated horizon evaluations at a cost orders of magnitude lower than CFD \cite{costa2023deep,sayghe2026fourier}. While we recognise differentiability as a particularly useful property of FNOs, this work does not exploit it, as explained later.

The central idea of this effort is to couple an FNO surrogate with MPC to enable closed-loop regulation of multiphase dynamics. We train an FNO to forecast the future evolution of a phase-indicator field (here, the volume fraction $\alpha$) over a finite horizon using a short history of recently saved states and a candidate control signal (here, inlet velocity). We then map the predicted fields to the controlled scalar observable (liquid level). The controller then selects an admissible actuation that tracks a time-varying setpoint schedule while enforcing hard bounds and penalising large control moves.
The algorithm is tested on a bubble column reactor case in a problem that involves controlling the reactor level. It is worth noting that the current configuration has some limitations, such as the fact that, as mentioned by Bieker et al. \cite{bieker2020deep}, most industrial configurations do not provide access to a complete domain description, as measurements are usually localised and discontinuous. A possible evolution of this approach might be the adoption of Physics-Informed Neural Operators, as initially introduced by \cite{li2024physics}, or other approaches that allow working with partial data.
\section{Numerical Model}
\paragraph{Problem formulation} The dynamics we intend to control are generally modelled with systems of partial differential equations. In the application described below, we solve a compressible multiphase flow problem in which the two phases are solved independently, and the interface is tracked by transporting the volume fraction $\alpha_i(\mathbf{x},t)$. Rusche and other authors introduced the Eulerian-Eulerian multi-fluid formulation  \cite{rusche2002computational,oliveira2003numerical} later incorporated it into OpenFOAM and named: \textit{twoPhaseEulerFoam}. We used the OpenFOAM solution algorithm to generate the training dataset and to simulate the controlled system.  
As anticipated, the phase volume fraction $\alpha_i$ is transported with the following: 
\begin{equation}
\frac{\partial \alpha_i}{\partial t}
+\nabla\cdot\!\left(\overline{\mathbf{U}}\,\alpha_i\right)
+\nabla\cdot\!\left(\mathbf{U}_r\,\alpha_i(1-\alpha_i)\right)=0,
\qquad i=1,2 .
\end{equation}
In which the time evolution of $\alpha_i$ depends on an advective term and an interface-related transport contribution that is non-zero only around the interface while vanishing in the pure phase regions. The mixture velocity, weighted on the volume fraction, and the relative velocity are defined as:
\begin{equation}
\overline{\mathbf{U}}=\alpha_1\mathbf{U}_1+\alpha_2\mathbf{U}_2,
\qquad
\mathbf{U}_r=\mathbf{U}_1-\mathbf{U}_2
\end{equation}
respectively. The phase-wise Eulerian momentum equations are:
\begin{equation}
\frac{\partial}{\partial t}\!\left(\alpha_i\rho_i\mathbf{U}_i\right)
+\nabla\cdot\!\left(\alpha_i\rho_i\,\mathbf{U}_i\otimes\mathbf{U}_i\right)
+\nabla\cdot\!\left(\alpha_i\rho_i\,\mathbf{R}_i^{\,\mathrm{eff}}\right)
= -\alpha_i\nabla p+\alpha_i\rho_i\mathbf{g}+\mathbf{M}_i,
\qquad i=1,2 .
\end{equation}
where $\rho_i$ and $\mathbf{U}_i$ are the density and velocity of phase $i$, $\partial(\alpha_i\rho_i\mathbf{U}_i)/\partial t$ is the accumulation of phase momentum, $\nabla\cdot(\alpha_i\rho_i\,\mathbf{U}_i\otimes\mathbf{U}_i)$ is convective transport of momentum with $\otimes$ denoting the dyadic (outer) product, $\nabla\cdot(\alpha_i\rho_i\,\mathbf{R}_i^{\mathrm{eff}})$ represents the divergence of the effective stress (viscous plus turbulent/Reynolds stress), $-\alpha_i\nabla p$ is the pressure-gradient force using a shared pressure field $p$, $\alpha_i\rho_i\mathbf{g}$ is the gravitational body force, and $\mathbf{M}_i$ is the interphase momentum transfer (e.g., drag and other coupling forces) providing two-way coupling between phases. The effective stress closure is
\begin{equation}
\mathbf{R}_i^{\,\mathrm{eff}}
=
-\nu_i^{\,\mathrm{eff}}
\left[
\nabla\mathbf{U}_i+(\nabla\mathbf{U}_i)^{T}
-\frac{2}{3}\mathbf{I}\,(\nabla\cdot\mathbf{U}_i)
\right]
+\frac{2}{3}\mathbf{I}\,k_i,
\qquad
\nu_i^{\,\mathrm{eff}}=\nu_i+\nu_i^{t}.
\end{equation}
with $\nu_i^{\mathrm{eff}}$ the effective kinematic viscosity (molecular $\nu_i$ plus turbulent/eddy viscosity $\nu_i^{t}$), $\nabla\mathbf{U}_i$ the velocity-gradient tensor, $(\nabla\mathbf{U}_i)^T$ its transpose, $\mathbf{I}$ the identity tensor, and $\nabla\cdot\mathbf{U}_i$ the velocity divergence; the bracketed term corresponds to the traceless strain-rate form for a Newtonian stress model. In the remaining term $k_i$ is the turbulent kinetic energy of the $i$-th phase whose isotropic contribution is included in the normal stresses.
\paragraph{Solution methodology}
As anticipated, the solver uses an Euler-Euler two-fluid formulation, treating both phases as continua. The volume fraction continuity equation is solved using the Multidimensional Universal Limiter with Explicit Solution (MULES) algorithm, which helps maintain the volume fraction within the bounds 0 and 1. The momentum equation is coupled with a Poisson-like pressure equation in the so-called "PIMPLE" \cite{weller1998tensorial} algorithm, which is a combination of Pressure-Implicit Splitting of Operators (PISO) \cite{issa1986solution} and Semi-Implicit Method for Pressure-Linked Equations (SIMPLE) \cite{patankar1981calculation}. 

% ============================================================
%  Neural Operators & FNO background
% ============================================================

\paragraph{Neural Operators}
Neural Operators (NOs) are a class of operator-learning models that approximate mappings between infinite-dimensional function spaces \cite{li2020fourier}:
\begin{equation}
    \mathcal{G}: \mathcal{A}(\Omega;\mathbb{R}^{m}) \rightarrow
                 \mathcal{W}(\Omega;\mathbb{R}^{n}),
\end{equation} where $\mathcal{A}(\Omega;\mathbb{R}^{m})$ and $\mathcal{W}(\Omega;\mathbb{R}^{n})$ are both Banach spaces, and $\mathcal{G}$ is a nonlinear operator mapping input
functions $a\in\mathcal{A}$ to output functions $w\in\mathcal{W}$. In the context of fluid dynamics, $\Omega\subset\mathbb{R}^d$ is the domain with
$d\in\mathbb{N}$; the input and output vector spaces $\mathbb{R}^{m}$ and $\mathbb{R}^{n}$ may have different dimensions. In this setting, $a(x)$ may represent an input field such as a coefficient, boundary condition, or source term, while $w(x)$ is the corresponding output, typically the solution of a system of PDEs. Among NOs, the Fourier Neural Operator (FNO) is particularly suitable for modelling fluid-dynamics problems, since fluid fields usually exhibit spatial correlations that are well represented in the Fourier basis. FNOs learn an approximation $\widehat{\mathcal{G}}_\theta\approx\mathcal{G}$ by composing layers that act globally across the entire domain $\Omega$, making it efficient at capturing long-range spatial dependencies. In the definition of $\widehat{\mathcal{G}}_\theta$, $\theta$ denotes the trainable parameters. 

% ============================================================
\paragraph{Fourier Neural Operators}
The methodology proposed by Li et al.\ \cite{li2020fourier} begins by lifting the input function $a(x)$ into a higher-dimensional latent representation:
\begin{equation}
    v_0(x) = P\!\left(a(x)\right),
\end{equation}
where $a(x)\colon\Omega\rightarrow\mathbb{R}^{m}$ is the input function and $P:\mathbb{R}^{m}\rightarrow\mathbb{R}^{d_v}$ is a learned local map that projects the input into the latent space $\mathbb{R}^{d_v}$, enabling the network to capture more complex structures. The output $v_0(x)\in\mathbb{R}^{d_v}$ serves as input to the subsequent operator layers, defined as:
\begin{equation}
    v_{l+1}(x) := \sigma\!\left(
        W_l\,v_l(x) + \left(\mathcal{K}_l\,v_l\right)(x)
    \right),
    \qquad \forall\,x\in\Omega,
\end{equation}
where $v_l(x)$ is the feature at location $x$ in the $l$-th layer, $l=0,\ldots,L-1$, $W_l:\mathbb{R}^{d_v}\rightarrow\mathbb{R}^{d_v}$ is a learned local linear transformation, and $\sigma$ is a nonlinear activation function applied componentwise to vectors in $\mathbb{R}^{d_v}$.
The nonlocal operator $\mathcal{K}_l$ is defined by:
\begin{equation}
    (\mathcal{K}_l v_l)(x)
    =
    \mathcal{F}^{-1}\!\left(
        R_l(k)\,\hat v_l(k)
    \right)(x),
    \qquad
    \hat v_l=\mathcal{F}(v_l),
\end{equation}
where \(k\in\mathbb{Z}^d\) is a discrete multi-index of Fourier modes. In practice, \(\mathcal{F}\) and \(\mathcal{F}^{-1}\) are implemented by the FFT on the computational grid, and only a finite set of low modes is retained:
\begin{equation}
    \mathcal{M}_l
    :=
    \left\{
        k=(k_1,\ldots,k_d)\in\mathbb{Z}^d
        :
        |k_j|\le M_{l,j},
        \quad j=1,\ldots,d
    \right\},
\end{equation}
with other modes \(k\notin\mathcal{M}_l\) set to zero.
For a \(d_v\)-channel latent field, the spectral multiplication is
\begin{equation}
    (R_l\hat v_l)_h(k)
    =
    \sum_{j=1}^{d_v}
    R_{l,hj}(k)\,\hat v_{l,j}(k),
    \qquad
    h=1,\ldots,d_v,
    \quad
    k\in\mathcal{M}_l,
\end{equation}
where \(R_l(k)\in\mathbb{C}^{d_v\times d_v}\). The \(l\)-th FNO layer is therefore written as:
\begin{equation}
    v_{l+1}(x)
    =
    \sigma\!\left(
        W_l\,v_l(x)
        +
        \mathcal{F}^{-1}\!\left(
            R_l(k)\,\hat v_l(k)
        \right)(x)
        +
        b_l(x)
    \right),     \qquad
    l=0,\ldots,L-1.
\end{equation}
where \(b_l:\Omega\to\mathbb{R}^{d_v}\) is an implementation-specific bias field. The final latent representation is projected to the output by
\begin{equation}
    w(x)=Q(v_L(x)),
\end{equation}
where \(Q:\mathbb{R}^{d_v}\to\mathbb{R}^{n}\) is a learned local transformation; in our implementation, \(Q\) is linear.
The model parameters \(\theta\) are trained by minimising the empirical loss:
\begin{equation}
    \mathcal{L}_{\mathrm{MSE}}(\theta)
    =
    \frac{1}{N}\sum_{i=1}^{N}
    \frac{1}{n\,N_yN_x}
    \left\|
        \widehat{Y}_i - Y_i
    \right\|_F^2,
\end{equation}
In the NMPC application described below, the generic input \(a\) is identified with an \(m\)-channel spatial field whose channels encode the state history, the coordinate channels, and the planned control sequence \(\bar u_k\). On the computational grid, this field is represented as a channel-stacked tensor. The generic output \(w\) is identified with the predicted volume-fraction trajectory over the horizon.
In this setting, temporal context is incorporated through the channel dimension rather than by enlarging the spatial domain: \(\Omega\) remains purely spatial, while both the \(K\) past snapshots and the \(H\) planned control moves are encoded through channels. This yields \(H\,d_u\) control channels in total. The generic output \(w\) comprises the \(H\) predicted volume fraction snapshots \(\{\alpha_{k+1},\ldots,\alpha_{k+H}\}\), so that
\begin{equation}
    \widehat{\mathcal{G}}_\theta\!\left(
        \alpha_{k-K+1:k},\,(x_x,x_y),\,\bar{u}_k
    \right)
    \approx
    \alpha_{k+1:k+H},
\end{equation}
with \(m = K + 2 + H d_u\) input channels and \(n = H\) output channels.

\paragraph{Neural Model Predictive Control}

In the following, we propose a neural model predictive control (NMPC) algorithm, in which the trained FNO surrogate \(\widehat{\mathcal{G}}_{\theta}\) is used to predict the future evolution of the system and Bayesian optimization (BO) is used to select the control action minimizing a finite-horizon objective. Let \(T_s>0\) denote the sampling interval and \(H\in\mathbb{Z}_{>0}\) the prediction horizon. The plant is the PDE system introduced above, which generates the volume fraction field \(\alpha(x,t)\) on the spatial domain \(\Omega\subset\mathbb{R}^d\), with spatial coordinate \(x\in\Omega\), under the action of a scalar input \(u(t)\in\mathcal U:=[u_{\min},u_{\max}]\). The control is applied with zero-order hold,
\begin{equation}
u(t)=u_k,\qquad t\in[kT_s,(k+1)T_s),
\end{equation}
where \(u_k\in\mathcal U\) is the input applied at step \(k\). Let the measured field at time step \(k\) be
\begin{equation}
\alpha_k(x):=\alpha(x,kT_s),
\end{equation}
and define the surrogate conditioning variable as the stack of the \(K\) most recent snapshots,
\begin{equation}
z_k:=\mathrm{col}\!\left(\alpha_{k-K+1},\ldots,\alpha_k\right)
\in [L^2(\Omega)]^K.
\end{equation}
The controlled output, is extracted from the volume fraction field through a functional:
\begin{equation}
y_k:=\ell(\alpha_k)\in\mathbb{R}.
\end{equation}
Let \(r_k\in\mathbb{R}\) denote the reference value at time step \(k\). At each control step, the FNO surrogate receives as input the state history \(z_k\), the spatial coordinate channels \((x_x,x_y)\), and a planned input sequence over the prediction horizon,
\begin{equation}
\bar u_k:=\mathrm{col}\!\left(u_{k|k},u_{k+1|k},\ldots,u_{k+H-1|k}\right)\in\mathcal U^H,
\end{equation}
and returns the predicted future fields
\begin{equation}
\widehat{\mathcal{G}}_{\theta}:\;
\left(z_k,\,(x_x,x_y),\,\bar u_k\right)
\mapsto
\widehat{\boldsymbol{\alpha}}_{k+1:k+H}
:=
\mathrm{col}\!\left(\widehat{\alpha}_{1|k},\ldots,\widehat{\alpha}_{H|k}\right),
\qquad
\widehat{\alpha}_{i|k}\approx \alpha_{k+i},
\end{equation}
where the relative index $i|k$ denotes the $i$-step-ahead prediction made at step $k$. In our implementation, the control sequence is embedded as boundary-supported input channels, so that the control acts only on the inlet boundary \(\partial\Omega_{\mathrm{in}}\). The corresponding predicted output trajectory is
\begin{equation}
\widehat y_{i|k}:=\ell\!\left(\widehat\alpha_{i|k}\right),
% ISSUE: You may want to state explicitly here that \(i=1,\ldots,H\), rather than leaving the range implicit.
\qquad
\widehat{\bar y}_k:=
\mathrm{col}\!\left(\widehat y_{1|k},\ldots,\widehat y_{H|k}\right).
\end{equation}
Because only a single scalar inlet input is optimized, we parameterize the planned sequence as constant over the horizon,
\begin{equation}
u_{k+i|k}=u_k,\qquad i=0,\ldots,H-1,
\end{equation}
so that
\begin{equation}
\bar u_k=u_k\,\mathbf 1_H\in\mathcal U^H.
\end{equation}
The optimal control problem therefore reduces to a one-dimensional search over \(u_k\in\mathcal U\). At each time step \(k\), we solve
\begin{align}
\min_{u_k\in\mathcal U}\quad
&J_k(u_k):=
\frac{1}{H}\sum_{i=1}^{H}\left(\widehat y_{i|k}-r_k\right)^2
% ISSUE: Using \(r_k\) for all \(i\) implicitly assumes the reference is frozen over the whole prediction horizon.
% ISSUE: If that is intended, state it explicitly; otherwise use something like \(r_{k+i|k}\).
+\lambda\left(u_k-u_{k-1}\right)^2
% ISSUE: \(u_{k-1}\) is undefined at the first control step unless you specify an initialization convention.
\label{eq:fno_mpc_obj}\\
\text{s.t.}\quad
&\widehat{\boldsymbol{\alpha}}_{k+1:k+H}
=
\widehat{\mathcal G}_{\theta}\!\left(z_k,\,(x_x,x_y),\,u_k\mathbf 1_H\right),
\label{eq:fno_mpc_pred}\\
&\widehat y_{i|k}=\ell\!\left(\widehat\alpha_{i|k}\right),
\qquad i=1,\ldots,H,
\label{eq:fno_mpc_output}\\
&u_{\min}\le u_k\le u_{\max},
\label{eq:fno_mpc_u_bounds}
\end{align}
where \(\lambda\ge 0\) penalizes aggressive input variations.
We solve \eqref{eq:fno_mpc_obj} using BO because the level functional \(\ell\) is discontinuous in the application considered below: specifically, the liquid level is obtained from \(\alpha\) via a thresholding operation, which renders the resulting objective not suitable for gradient-based optimisation. At each BO iteration, a Gaussian-process surrogate is fitted to the set of previously evaluated pairs
\begin{equation}
\mathcal D_k=\left\{\left(u^{(i)},J_k(u^{(i)})\right)\right\}_{i=1}^{n},
\end{equation}
% ISSUE: \(n\) is introduced without being defined as the current BO iteration count or current dataset size.
yielding a posterior mean \(m_k(u)\) and variance \(s_k^2(u)\). An acquisition function, here chosen as expected improvement (EI), is then maximized to balance exploitation of low-cost regions and exploration of uncertain ones:
\begin{equation}
u^{(n+1)}=\arg\max_{u\in\mathcal U}\mathrm{EI}_k(u).
\end{equation}
After a fixed number of BO iterations, we apply the minimizer \(u_k^\star\)  to the plant, measure the new field snapshot \(\alpha_{k+1}\) and update the history \(z_{k+1}\)  before repeating the procedure.
Finally, when \(\ell\) is smooth, gradient-based optimization is more appropriate. Indeed, \(\widehat{\mathcal G}_{\theta}\) is fully differentiable, as discussed in~\cite{cheng2025accelerating}, and the gradient  \(dJ_k/du_k\) could in principle be computed by automatic differentiation.
In the present setting, however, BO remains particularly attractive because the decision variable is scalar, thereby avoiding the curse of dimensionality that would otherwise limit its practicality in higher-dimensional MPC parameterisations.

\section{Application: Bubble Column Reactor Control}

\begin{figure}
    \centering
    \includegraphics[width=1.\linewidth]{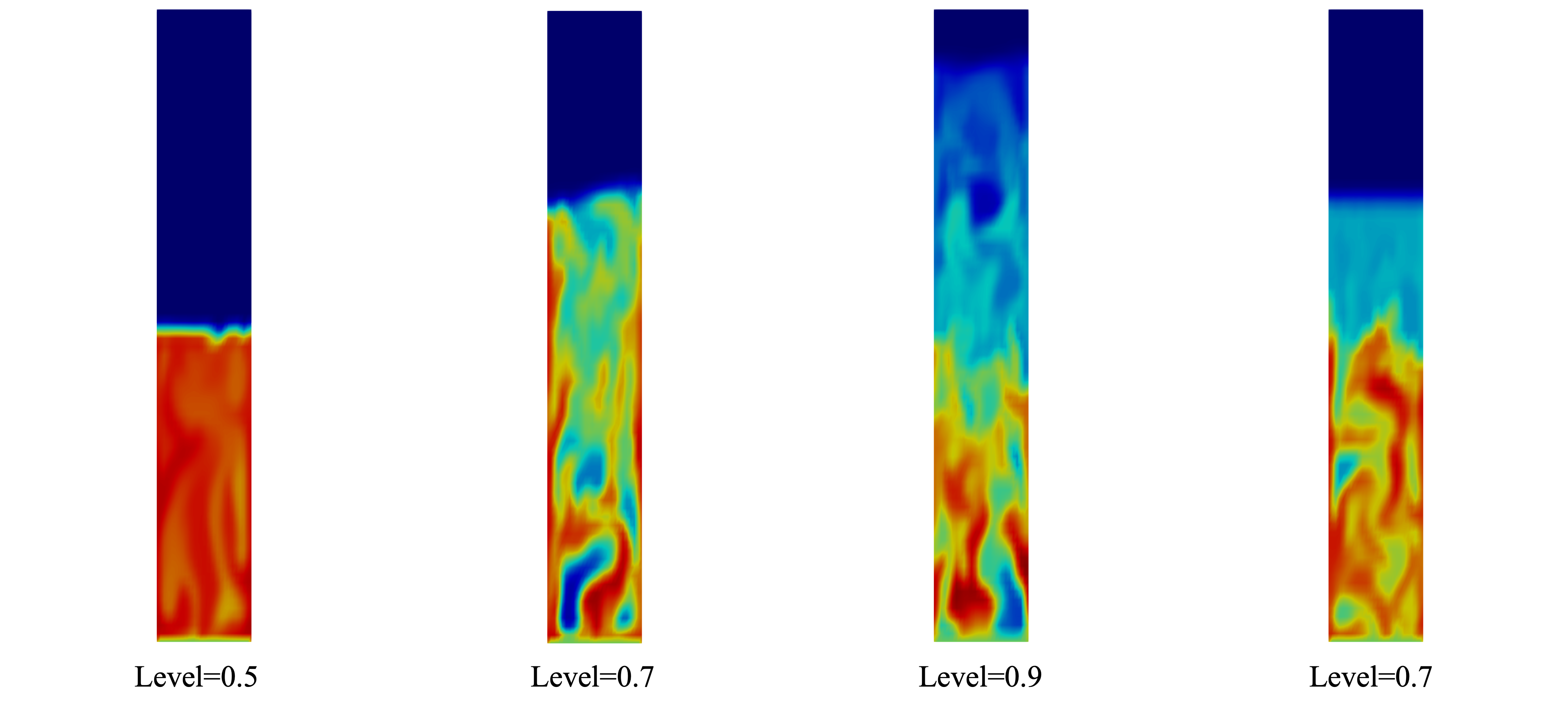}
\caption{Representative snapshots of the bubble-column phase field (volume fraction) under closed-loop operation, shown for different commanded liquid levels. The interface location and the internal recirculation/bubble-plume structures evolve nonlinearly with the inlet actuation, highlighting the strongly state-dependent hydrodynamics captured by the CFD dataset.}
    \label{fig:1}
\end{figure}

The framework detailed in the previous section is applied to the problem of controlling a bubble column reactor. The above consists of a system in which a column of liquid is put in motion by the injection of a gas introduced from the bottom (\ref{fig:1}). These reactors are designed to favor heat and mass transfer between the gas and liquid phases by generating an extremely large interfacial area through the formation of small bubbles, usually through a sparger. All this complexity can only be accurately handled by sophisticated CFD models, since most diagnostics in real-life scenarios lack spatial or temporal resolution \cite{yan2024experimental}.  
Designing a control system for such devices has therefore always been challenging, and the most common approach has been to either adopt passive control mechanisms or build control on sparse data or, eventually, on reduced-order models. In the following paragraphs, we detail the computational domain and simulation parameters, the generated dataset information, and the training results. Finally, the control performance is highlighted. 

\paragraph{Computational domain and mesh}
The computational domain consists of a single structured hexahedral block. The domain spans $L_x=0.15~\mathrm{m}$, $L_y=1.0~\mathrm{m}$, and a finite out-of-plane thickness $L_z=0.10~\mathrm{m}$ (extruded mesh), with vertices defined at
$(0,0,0)$, $(0.15,0,0)$, $(0.15,1,0)$, $(0,1,0)$ and their counterparts at $z=0.10~\mathrm{m}$. The block is discretized into $(N_x,N_y,N_z)=(25,75,1)$ cells with uniform grading. The corresponding uniform cell spacings are $\Delta x = L_x/N_x = 6.0\times 10^{-3}~\mathrm{m}$, $\Delta y = L_y/N_y \approx 1.33\times 10^{-2}~\mathrm{m}$, and $\Delta z = L_z/N_z = 0.10~\mathrm{m}$.
\paragraph{Initial and boundary conditions (two-phase fields)}
The simulations were initialized with a stratified phase distribution using the volume fraction of the gas phase, $\alpha_\mathrm{air}$. A total of $750$ cells were initialized with $\alpha_\mathrm{air}=0$ and $1125$ cells with $\alpha_\mathrm{air}=1$, corresponding to an interface located at $y \approx (30/75)\,L_y \approx 0.40~\mathrm{m}$ (water below, air above).  At the inlet we set $\alpha_\mathrm{air}$ to 1 (pure air), and a step function for the velocity. No-slip conditions ($\mathbf{U}=\mathbf{0}$) were enforced on the side walls for both phases.

\paragraph{Dataset generation}

The dataset was created by randomly generating cases with different input velocities changing over discrete time intervals. The evolution of the system was captured every \(0.1\,\mathrm{s}\) for a total of \(100\) steps over \(10\) seconds. For each run, we stored only the volume fraction time series \texttt{alpha} with shape \((T,N_y,N_x)\) and the corresponding inlet-velocity signal with shape \((T,C)\) (or \((T,)\) for scalar control, reshaped to \((T,1)\)), where \(C\) is the control dimension. We then segmented these time series into fixed-length temporal sliding windows to form the learning pairs used for training. The input portion consists of the \(K\) most recent volume fraction fields, \(\alpha_{t-K+1:t}\), together with the planned inlet sequence over the next \(H\) control intervals, \(u_{t:t+H-1}\). The target is the future volume fraction trajectory \(\alpha_{t+1:t+H}\).

\begin{figure}
    \centering
    \includegraphics[width=\linewidth]{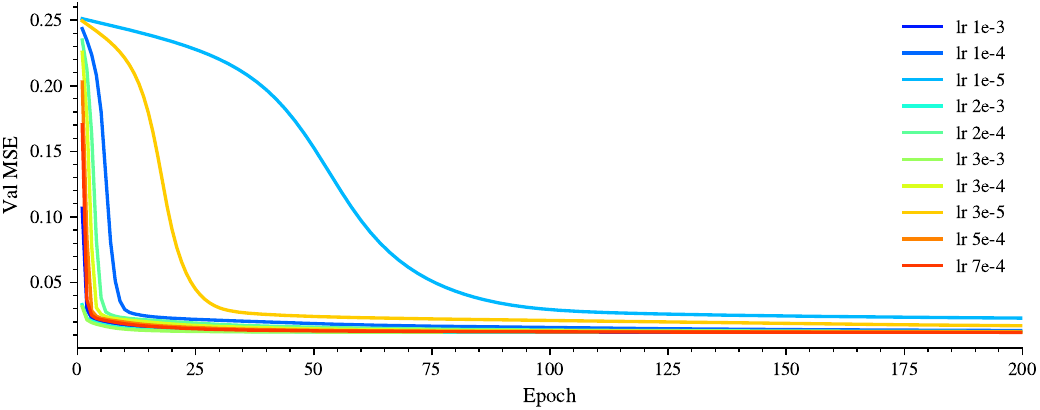}
\caption{Validation mean squared error (MSE) as a function of training epoch for different learning rates. All configurations show a rapid initial drop in error followed by gradual convergence toward a similar low-MSE plateau. Larger learning rates accelerate early convergence, while smaller learning rates reduce the error more gradually; overall, learning rates in the intermediate range provide the best balance between convergence speed and final validation performance.}
    \label{fig:2}
\end{figure}
To match the operator-learning architecture, we encode the network input as a channel-stacked tensor $X_t\in\mathbb{R}^{C_{\mathrm{in}}\times N_y\times N_x}$  with $C_{\mathrm{in}}=K+2+HC$, where the additional $2$ accounts for the 
spatial coordinate channels $(x_x,x_y)$ appended point-wise to the input. The first $K$ channels are the previous fields. The control is encoded by reshaping the future sequence into $HC$ channels and injecting it only at the inlet through a fixed spatial mask $M\in\{0,1\}^{N_y\times N_x}$ (in our implementation, $M=1$ on the bottom row corresponding to the inlet boundary $\partial\Omega_{\mathrm{in}}$ at $y=0$, and $0$ elsewhere). The above produced control channels of the form
\begin{equation}
    U_t^{(h,i,j)} = \tilde{u}_t^{(h)} \cdot M_{i,j},
    \qquad
    h = 1,\ldots,HC,\quad
    (i,j)\in\{1,\ldots,N_y\}\times\{1,\ldots,N_x\},
\end{equation} where $\tilde{u}_t \in \mathbb{R}^{HC}$ is the planned control sequence $u_{t:t+H-1}$ flattened into $HC$ scalars, and $M \in \{0,1\}^{N_y \times N_x}$ is the inlet mask defined above. These control channels are concatenated with the history and coordinate channels to form the full input tensor
\begin{equation}
    X_t = \bigl[\,\alpha_{t-K+1:t}\ \big|\ (x_x,x_y)\ \big|\ U_t\,\bigr]
    \in \mathbb{R}^{(K+2+HC)\times N_y\times N_x}.
\end{equation}
The training target was:
\begin{equation}
    Y_t = \alpha_{t+1:t+H} \in \mathbb{R}^{H\times N_y\times N_x}.
\end{equation}

\paragraph{Training}

\begin{figure}
    \centering
    \includegraphics[width=\linewidth]{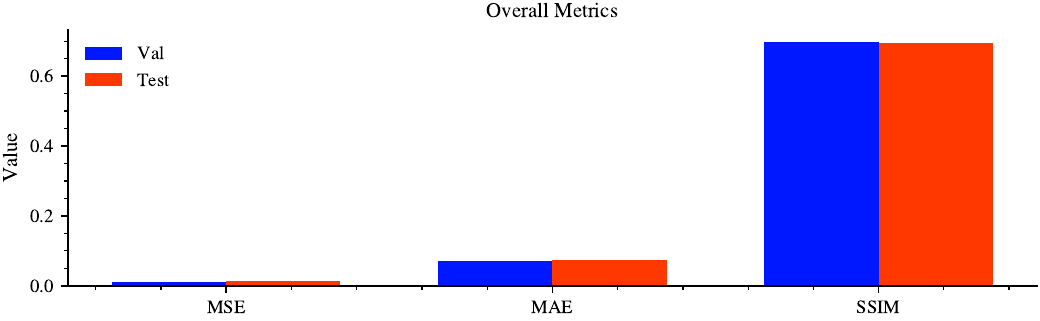}
    \caption{Comparison of overall prediction accuracy on the validation and test sets using mean squared error (MSE), mean absolute error (MAE), and structural similarity index (SSIM). The similar values across both splits suggest stable performance and good generalization of the forecasting model.}
    \label{fig:3}
\end{figure}

\begin{figure}
    \centering
    \includegraphics[width=1.\linewidth]{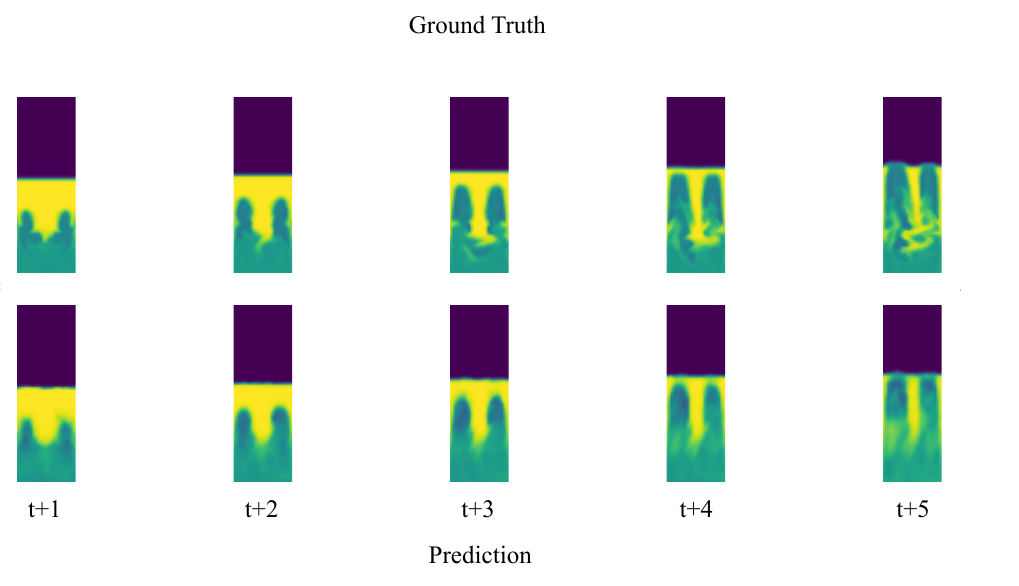}
\caption{Volume fraction field forecasting over a 5-step horizon with the trained Fourier Neural Operator (FNO). Top row: ground-truth volume fraction fields at $t\!+\!1$ to $t\!+\!5$ from the CFD model. Bottom row: corresponding FNO predictions, demonstrating the surrogate’s ability to propagate interface motion and large-scale structures.}
\label{fig:4}
\end{figure}

We trained a Fourier Neural Operator (FNO) on the resulting windowed dataset to learn an operator that maps a short history of volume fraction fields and the planned inlet-velocity profile to future volume fraction snapshots. Training was executed on a single NVIDIA Quadro GV100 (32 GB) with PyTorch 2.10.0, CUDA 12.2, and driver 535.288.01. The final optimised configuration used a batch size of 512 with GPU-resident preloading and CuPy preprocessing \cite{okuta2017cupy,paszke2019pytorch}. We performed tests at various learning rates achieving satisfactory results in a few epochs for a learning rate of 2e-4 \ref{fig:2},\ref{fig:3} . 
As illustrated in Fig.~\ref{fig:4}, the learned operator reproduces the global interface dynamics reliably; however, it tends to under-resolve fine-scale structures within the liquid phase, leading to small discrepancies in the detailed volume fraction distribution.

\paragraph{Training acceleration benchmarks}

We benchmarked training on the same V100 system across multiple batch sizes and optimization settings. The best configuration achieved a 3.09x speedup over the single-GPU baseline at a batch size of 512. Pre-loading the dataset to GPU memory incurred a one-time cost of 4.3 ms per sample (about 50 s for 11,900 samples) and reduced data-transfer overhead during training. Inference throughput peaks at batch size 256, with 0.041--0.051 ms per sample (19,711--24,508 samples/s).

\begin{table}[h!]
\centering
\small
\begin{tabular}{l r r r}
\hline
\textbf{Configuration} & \textbf{Batch} & \textbf{Time/Epoch (s)} & \textbf{Speedup} \\
\hline
Baseline (single-GPU) & 64 & 12.13 & 1.00x \\
Optimized (preload+CuPy) & 64 & 5.73 & 2.12x \\
Baseline (single-GPU) & 256 & 13.34 & 1.00x \\
Optimized (preload+CuPy) & 256 & 5.08 & 2.62x \\
Baseline (single-GPU) & 512 & 16.96 & 1.00x \\
Optimized (preload+CuPy) & 512 & 5.48 & 3.09x \\
\hline \\
\end{tabular}
\caption{Training time per epoch on a single NVIDIA Quadro GV100. The optimized configuration uses GPU-resident preloading and CuPy preprocessing.}
\label{tab:training_speedup}
\end{table}

\begin{figure}
    \centering
    \includegraphics[width=\linewidth]{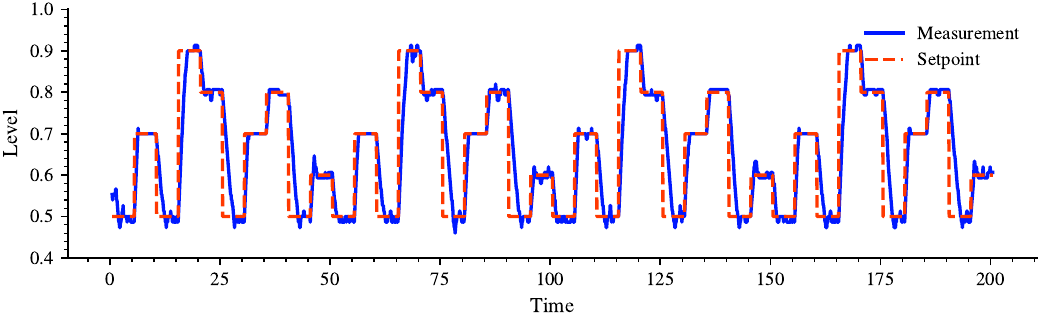}
\caption{Closed-loop setpoint tracking performance for the bubble-column liquid level. The measured level (blue) is regulated to a piecewise-constant reference (red dashed) across repeated setpoint changes, with short transients at switching times and small steady-state offsets near the operating extremes.}
    \label{fig:5}
\end{figure}
\begin{figure}
    \centering
    \includegraphics[width=\linewidth]{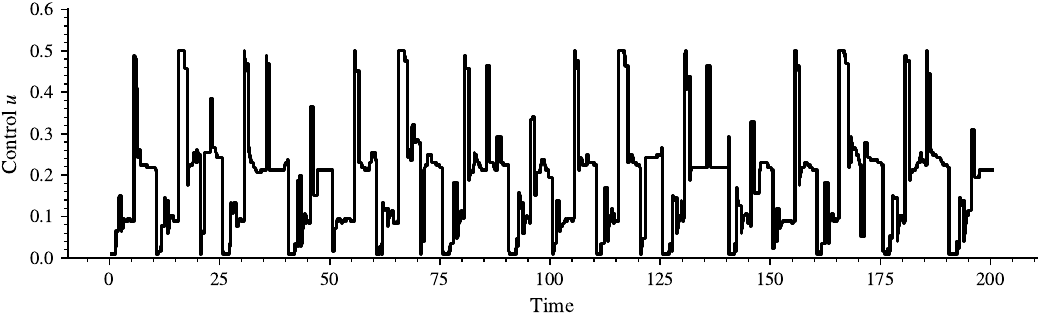}
\caption{Control action applied by the surrogate-assisted MPC. The inlet velocity is implemented as a constant-hold input updated at each control interval; the optimizer selects bounded actuation levels that balance tracking error reduction with conservative input usage over the full experiment horizon.}
    \label{fig:6}
\end{figure}

\begin{figure}[!ht]
\centering
\begin{subfigure}[b]{\textwidth}
  \includegraphics[width=1\linewidth]{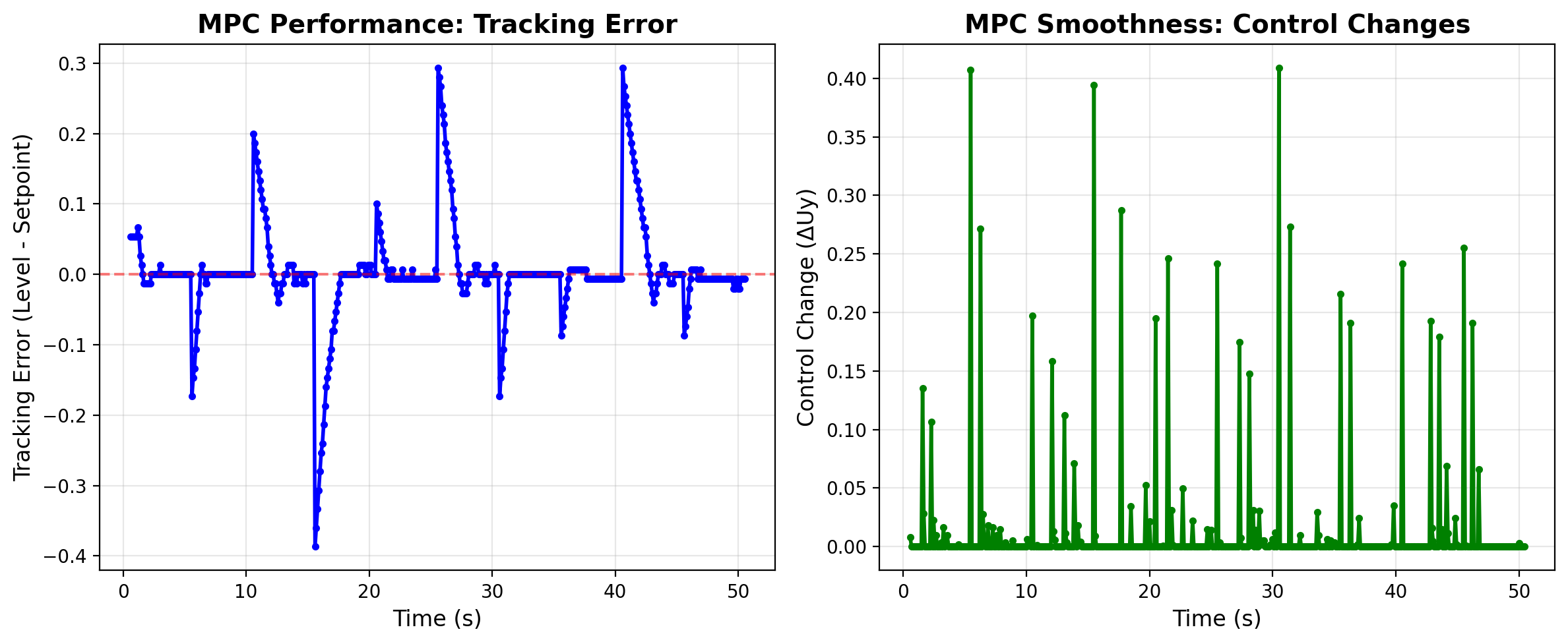}
  \caption{}
  \label{fig:7} 
\end{subfigure}
\medskip
\begin{subfigure}[b]{\textwidth}
  \includegraphics[width=1\linewidth]{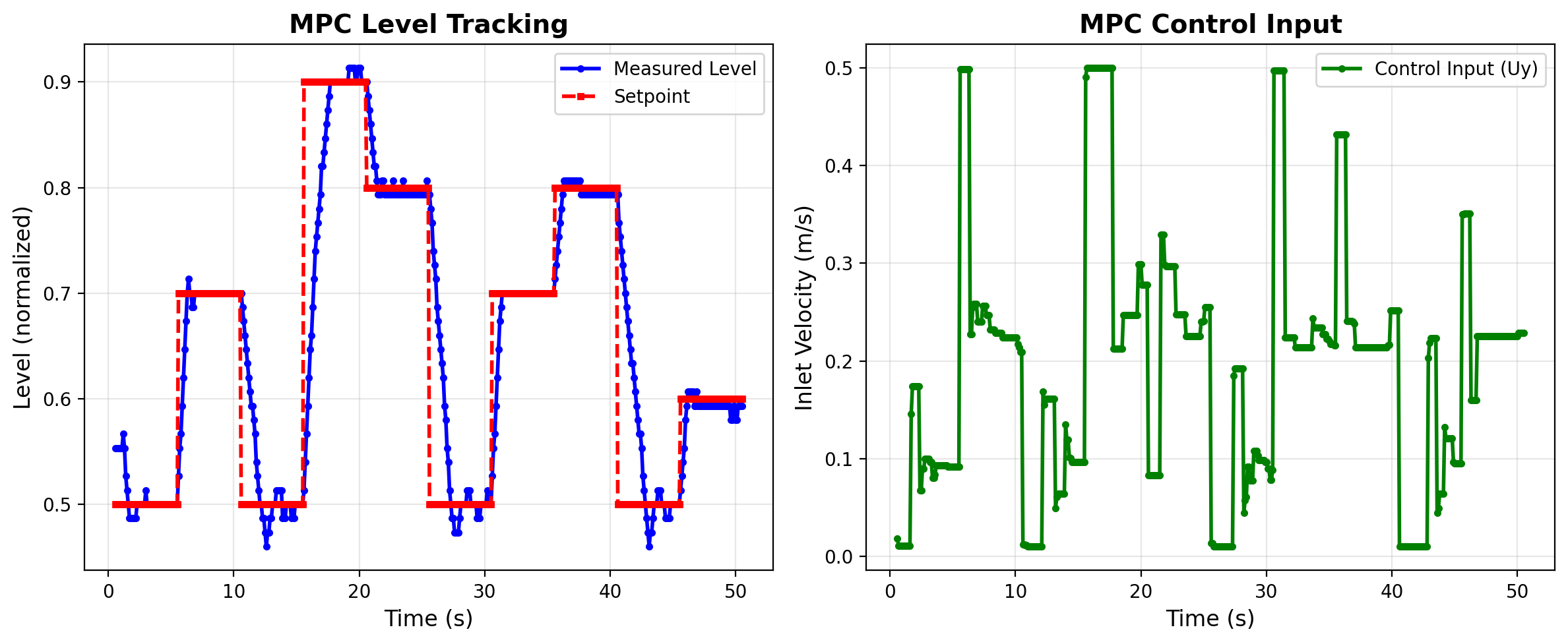}
  \caption{}
  \label{fig:8}
\end{subfigure}

\caption{MPC performance for liquid-level control under a time-varying setpoint. (a) Top row: level tracking (left) and control input \(u\) (right). The measured level follows the setpoint closely, with only brief overshoot and undershoot during transitions, while the inlet-velocity control action remains bounded. (b) Bottom row: tracking error (left) and control increments \(\Delta u\) (right). The tracking error remains close to zero for most of the simulation, with larger deviations occurring mainly at setpoint changes, while the control increments indicate generally smooth actuator behavior with occasional sharp corrections.}
\end{figure}

\paragraph{Tracking quality and transient behavior}
Figure~\ref{fig:5} compares the measured liquid level, in blue, against the reference setpoint, in red, piecewise constant. Overall, the controller achieves consistent tracking across repeated setpoint changes while exhibiting only short transients at switching times and is equilibrated by sudden changes in gas flowrate as shown in Fig. \ref{fig:6}. The response is well-damped: after each step change, the level converges rapidly toward the new target without sustained oscillations, indicating that the closed-loop dynamics are stable over the entire test horizon. The most prominent errors occur immediately after a shift in the setpoint, when the system's characteristic response time prevents faster adaptation of the level as better highlighted in \ref{fig:7}.    

\paragraph{Steady-state offsets and constraint effects}
Once the short transition phase is excluded, the remaining steady tracking error is small for intermediate setpoints (e.g., $0.6$--$0.8$), while a systematic bias becomes visible at the extremes. In particular, at the highest commanded level ($\mathrm{setpoint}=0.9$) the response tends to settle slightly below the target (typical mean level $\approx 0.88$), whereas at the lowest commanded level ($\mathrm{setpoint}=0.5$) the response can remain marginally above the reference in some plateaus. This behavior is consistent with the input being bounded. Importantly, the offset remains bounded and repeatable, suggesting the controller is robust but operates close to its feasible actuation envelope at the extremes.

\section{Conclusions}

We introduced a framework for controlling complex multiphase phenomena that occur over short characteristic times. We applied a model predictive control strategy with neural operators for fast forecasting of complex dynamics exploiting Fourier Neural Operators' ability to accurately and promptly predict the spatio-temporal evolution of systems described by PDEs. The method's low latency is essential in many engineering applications, including a bubble column reactor, which is the case we chose to validate the method. 
We demonstrated the ability to adapt the gas inlet velocity to match a given level of gas-entraining liquid in the system, a feat that requires capturing the nonlinear relationship between the amount of gas that escapes the free surface and the amount introduced.  Rather than relying on reduced-order models tied to a fixed discretisation or operating point, the controller evaluates candidate actuations through field-level rollouts and maps the predicted states to a task-relevant observable, such as the liquid level in this case, using a receding-horizon objective with input bounds and regularisation. Across repeated piecewise-constant setpoint changes, we achieved consistent tracking with short transients at switching times and stable behaviour throughout the evaluation horizon, indicating that the surrogate-assisted MPC can regulate nonlinear multiphase dynamics without embedding high-fidelity CFD in the optimisation loop. Notably, the characteristic time scale of the computation is extremely small, ranging from 1e-3 to 1e-2 at each time step throughout the optimization process. On the Quadro GV100 system, FNO inference throughput reaches 19,711--24,508 samples/s (0.041--0.051 ms per sample) depending on batch size (Table \ref{tab:training_speedup}) .
Overall, the results support the use of learned operators as an enabler of real-time control while a direct wall-clock comparison against an embedded CFD-MPC implementation is left for future work. By forecasting the full field, the framework naturally generalises beyond scalar regulation. It provides a foundation for future extensions to partial observability \cite{lewis2010reinforcement,brunton2016discovering}, uncertainty-aware decision making \cite{bulte2025probabilistic}, and physics-informed operator learning \cite{boya2024physics,yamazaki2025finite}, which are expected to further improve robustness and applicability to industrial multiphase unit operations. We also want to note that, beyond speed, forecasting a field rather than a scalar output might enable more refined applications, such as controlling the distribution of the gas phase across particular structures (e.g., more or less active regions of a catalyst bed) in the system. While MPC ultimately evaluates performance through a task-relevant functional, an accurate prediction of that functional still requires capturing transport, accumulation, and interfacial motion over the whole domain. A field-level surrogate retains the spatial structure necessary to represent these mechanisms, while the output functional provides a consistent and robust bridge from predicted fields to control decisions. Moreover, the differentiability of the surrogate enables (in principle) gradient-based optimization. However, this possibility has not been explored in this work. 

\section*{Acknowledgements}
The authors sincerely thank the open-source communities that made this work possible. In particular, we acknowledge the developers and contributors of the \textbf{Python} ecosystem, and \textbf{OpenFOAM} community. This work used Python 3.11.14, PyTorch 2.10.0, CuPy 13.6.0, RAPIDS 25.12, CUDA 12.2, and NVIDIA driver 535.288.01 on a single Quadro GV100 (32 GB). Special recognition is given to the developers of \texttt{twoPhaseEulerFoam} and related two-phase solvers. This research was supported by \textbf{King Abdullah University of Science and Technology (KAUST)}. 

%Bibliography
\bibliographystyle{unsrt}  
\bibliography{references}  

\end{document}